\documentstyle[12pt]{article}
\newcommand{\clequ}{\setcounter{equation}{0}}
\begin{document}

\begin{titlepage}
\begin{center}
\ \\
\vspace{3.cm}
{\Large\bf The representations of the Hubbard algebra
in terms of spin-fermion operators and motion of a hole in
an antiferromagnetic state}

\vskip 3.cm
{\large  V.I.Belinicher and A.L.Chernyshev} \\
Institute of Semiconductor Physics, 630090 Novosibirsk, Russia\\
\end{center}
\vspace{2.cm}

\begin{abstract}
The representation of the Hubbard operators in terms of the spin$-\frac{1}{2}$
operators and the fermion operator with spin$-\frac{1}{2}$ is proposed.
  In the low-energy limit this representation is reduced to the
 representation following from the Hubbard diagramm technique.
In framework of this approach motion of a hole in an antiferromagnetic
state of the t-J model is considered. It is shown that the primary hole
energy is strongly renormalized and the band width has an order of J rather
than t.
  The functional integral for the strongly correlated model
induced by the obtained representation  is formulated. The
representation of the total Hubbard algebra for states in the lower
and the upper Hubbard bands is formulated in terms of the spin$-\frac{1}{2}$
and two fermion fields with spin$-\frac{1}{2}$ is formulated.


\end{abstract}

\vfill
\begin{center}
Novosibirsk\\
1992
\end{center}

\end{titlepage}

\section{ Introduction}
\clequ \
The strong correlation in the electron system can be described by the
restriction of the number of states at every lattice site.  Such
restriction can be described by the introduction of the Hubbard operators.
But the Hubbard operators are not convenient for description of the
statistical and dynamical properties of the strongly correlated electron
system.  That is why some representation of the Hubbard operators in
terms of more convenient  operators are used.  The well known
representation of such type is slave-boson (fermion)
representation \cite{ba,co,ca}. These representations are based on the
mapping of three electron states  $|0>, |\uparrow >, |\downarrow >$ at
a lattice site into fermion and boson Gilbert space and restriction of the
one-particle states.

This work has mainly methodical character.  At first the
simple historical review of the description of spin system on the basis of
the Holstein-Primakoff and the Dyson-Maleev   representation is presented.
Further the mapping of three Hubbard states into the eight spin$-\frac{1}{2}$
and fermions (with a spin$-\frac{1}{2}$) Gilbert space is produced.
Such mapping generates the representation of the Hubbard operators in terms of
fermion
and spin$-\frac{1}{2}$ operators.  Such representation can be useful in some
applications.

Renormalization of the hole energy in a antiferromagnetic state of the
 $t-J$ model is considered in the next section. This consideration is
based on the approach to an antiferromagnetic state developed in
work \cite{be}. We will show that the method developed by Kane, Lee and
Read \cite{ka} for the description of the motion of a single hole in
quantum antiferromagnet is also applicable to the our approach. We will show
that the hole spectrum is strongly renormalized in the limit $t \ll J$
where $t$ and $J$ are the hopping and exchange constants for the $t-J$
model. The quasiparticle band has the width $J$ and the quasiparticle
residue is equal to $J/t$.

After that two methods of obtaining the functional integral
for strongly correlated electron system is discussed.
  In final part of
the work the representation of the total Hubbard algebra is constructed
and its physical interpretation in terms of the local spins, electrons in
the upper Hubbard bands and holes in the lower Hubbard bands is presented.

\section{Spin systems}
\clequ \

For the clarification of the structure of representation for the Hubbard
model we want to remember connection between the Holstein-Primakoff (HP)
 \cite{hp}
and the Dyson-Maleev (DM) \cite{dy,ma} representations. They based on
mapping of the 2S+1 spin states into the lowest  2S+1
states of the Heisenberg algebra:
\begin{eqnarray}
\label{111a}
&&S^{-}_{HP} =  P(a^{+}a)\sqrt{2S-a^{+}a}\ a \ P(a^{+}a) \nonumber \\
&&S^{+}_{HP} =  P(a^{+}a)a^{+}\sqrt{2S-a^{+}a} \ P(a^{+}a)  \nonumber \\
&&S^{z}_{HP} =  P(a^{+}a)(-S+a^{+}a)P(a^{+}a)
\end{eqnarray}
here $ S^{+},S^{-}, S^{z} $ are spin operators, $ a^{+}, a $ are creation
and annihilation Bose operators,  $ P(a^{+}a) $ is the projector operator
on the lower  2S+1 states: $ |0>, |\uparrow >, ...|2S> $.  The
representation (\ref{111a}) is the direct operator identity: the matrix
elements of the original spin operator ${\bf S}$ and the HP spin operators
${\bf S}_{HP}$ are the same on the lowest  2S+1  states of the Heisenberg
algebra.
The DM representation for spin operators
has the form:
\begin{eqnarray}
\label{22a}
&&S^{-}_{DM} =  a \nonumber \\
&&S^{+}_{DM} =  a^{+}(2S-a^{+}a) \nonumber \\
&&S^{z}_{DM} =  -S+a^{+}a
\end{eqnarray}
This representation does not represent a direct operator identity
but the spin algebra is valid for DM spin operators (\ref{22a}) also.
One can check that there exists the operator $V(a^{+}a)$ with property
\begin{equation}
\label{3a}
{\bf S}_{HP}=PV^{-1}{\bf S}_{DM}VP
\end{equation}
here ${\bf S}_{HP}, {\bf S}_{DM}$ are the HP (\ref{111a}) and the
DM (\ref{22a}) spin operators. The identity (\ref{3a}) means that matrix
elements of both parts of Eq.(\ref{3a})  between any states $<n| $
and $| n^{`}>$ are equal to each other. If $|n,n^{`}>2S$ these matrix
elements are equal to zero. The explicit form of the
matrix elements $V$ is following
\begin{eqnarray}
\label{4a}
<n^{`}| V|n> =  \delta_{n^{`},n}\prod_{m=0}^{n}
\sqrt{2S+1-m},\ \ \ \    n,n^{`}\leq2S
\end{eqnarray}
If we want to calculate the partition function
\begin{equation}
\label{5a}
Z=Tr(\exp(-\beta H({\bf S}))=Tr(P\exp(-\beta H({\bf S}_{HP}))
\end{equation}
we can use the identity (\ref{3a}) and the commutation of the operators $V$
 and $P$. As a result, we have for Z
\begin{equation}
\label{6a}
Z=Tr(P\exp(-\beta H({\bf S}_{DM}))  .
\end{equation}
Eq.(\ref{6a}) is valid because the action of ${\bf S}_{DM}$
on the state $|n>$ for $n=0,1,...,2S$ leads only to the same states,
 i.e. the lower $2S+1$ states form an invariant subspace with respect
to the DM spin operators ${\bf S}_{DM}$.

If we are interested in the properties of the partition function (\ref{5a})
 at the low temperatures in the ferromagnetic or antiferromagnetic
state, we can   omit the projector $P$ in formula (\ref{6a}) because
the contribution to the higher states $|n>$ with $n>2S$ into the
trace (\ref{6a}) is
exponentially small over the parameter $\beta J$ where $J$ is an
exchange integral. This discussion explains the correctness
of using the DM representation for the description of the low-energy
processes at the low temperatures in ferromagnets and antiferromagnets.

\newpage

\section{Strongly correlated systems.}
\clequ \

A similar interpretation can be given for the representation for the
Hubbard operators $X^{ab}$. Let us consider the Gilbert space at
every lattice site representing the direct product of a Gilbert space
of a spin$-\frac{1}{2}$ and fermion with a spin$-\frac{1}{2}$.
The total number
of states in that spin-hole Gilbert space is equal to eight:
$|0,\sigma>$, $|1,\sigma^{`},\sigma>$, $|2,\sigma>$,
here the first index represents the number of fermions; $\sigma,
\sigma^{`}=\uparrow ,\downarrow$ are spin projections. One can
introduce the Fermi operators $h^{+}_{\sigma },h_{\sigma }$ then
\begin{equation}
\label{7a}
|1,\sigma^{`},\sigma>=h^{+}_{\sigma^{`}}|0,\sigma>, \ \ \
\ \ \ |0,\sigma>=h_{\sigma^{`}}|1,\sigma^{`},\sigma>.
\end{equation}

The singlet and triplet states can be formed from one-fermion states:
\begin{eqnarray}
\label{8a}
&& |s>=(1/\sqrt{2})(|1\uparrow,\downarrow>-|1\downarrow,\uparrow>),
\ \ \ \ \ |t1>=|1\uparrow,\uparrow> \nonumber \\
&& |t0>=(1/\sqrt{2})(|1\uparrow,\downarrow>+|1\downarrow,\uparrow>),
\ \ \ \ |t-1>=|1\downarrow,\downarrow>.
\end{eqnarray}
We can map the initial Gilbert space of the Hubbard model without
two-fermion  states into the spin-hole Gilbert space:
\begin{equation}
\label{9a}
|0>\Rightarrow|s>,\ \ \ \ \ \ |1,\sigma>\Rightarrow|0,\sigma>,
\end{equation}
and obtain the following representation of the Hubbard operators
$X^{ab}$ in terms of the holes  $h^{+}_{\sigma },h_{\sigma }$ and
the spin$-\frac{1}{2}$ operators {\bf s}:
\begin{eqnarray}
\label{99a}
&& X^{0\sigma}_{HP}=2\sigma\sqrt{2}(h^{+}\hat{S})_{-\sigma}(1-\hat{n})
\nonumber \\
&& X^{\sigma 0}_{HP}=2\sigma\sqrt{2}(1-\hat{n})(\hat{S}h)_{-\sigma }
\nonumber \\
&& \hat{N}_{HP}=X_{HP}^{\uparrow \uparrow}+X_{HP}^{\downarrow \downarrow}=
1-\hat{n}+\hat{d}
\nonumber \\
&& X_{HP}^{00}=(1/4)(\hat{n}-2\hat{d}-2(h^{+}{\bf \sigma}h){\bf s}) \nonumber
\\
&& {\bf S}={\bf s}(1-\hat{n}+\hat{d})
\end{eqnarray}
here
\begin{eqnarray}
\label{10a}
\hat{S}=(1/4)(1-2{\bf s}{\bf \sigma }),\ \ \
\hat{n}=(h^{+}h),\ \ \ \hat{d}=h^{+}_{\uparrow}h_{\uparrow}
h^{+}_{\downarrow}h_{\downarrow}.
\end{eqnarray}
This representation is the direct operator identity: (1) the matrix
elements $X_{HP}^{ab}$ operators (\ref{99a}) between the physical states $|s>,
\ |0 \sigma>$ are the same as for the initial Hubbard operators
$X^{ab}$; (2) the matrix elements between the unphysical states $|t
m>, |2 \sigma>$ are equal to zero; (3) the matrix elements between
physical states $|s>,|0 \sigma>$ and unphysical states
$|t m>, |2 \sigma>$  are equal to zero. The representation (\ref{99a})
for $X_{HP}^{ab}$ certainly is hermitian and does not require any
constraints. If we compare the representation (\ref{99a}) with initial
expression of the Hubbard operators $X^{ab}$ in terms of the physical
Fermi operators of electrons we can see that the number of electrons
is close to unity while the number of Fermi holes is small at
filling closed to unity. The empty space without electrons can
be imaged as the bound singlet state of the Fermi hole and the
spin$-\frac{1}{2}$ with the total spin equal to zero.

In addition to representation (\ref{99a}) in work  \cite{be}
was proposed the nonhermitian representation for the Hubbard operators
\begin{eqnarray}
\label{11a}
&& X_{DM}^{0\sigma}=2\sigma h^{+}(1/2-({\bf \sigma s})+(h^{+}h))_{-\sigma },
\ \ \ X_{DM}^{\sigma 0}=2\sigma h_{-\sigma} ,
\nonumber \\
&& \hat{N}_{DM}=X_{DM}^{\uparrow \uparrow}+
X_{DM}^{\downarrow \downarrow}=1-\hat{n}, \ \ \
X_{DM}^{00}=\hat{n}, \ \ \ {\bf S}={\bf s}+(1/2)(h^{+}{\bf \sigma}h).
\end{eqnarray}
This representation does not present direct operator identity but the
Hubbard algebra is valid for this representation.  The representation
(\ref{11a}) can be named as the Dyson-Maleev representation for the Hubbard
operators.

What is the connection between representations (\ref{99a}) and
(\ref{11a}) for the Hubbard operators ? One can check that if we produce
canonical transformation of the hole Fermi operators in (\ref{11a})
$h_{\sigma} \rightarrow \sqrt{2}h_{\sigma},\  h^{+}_{\sigma} \rightarrow
\  (1/\sqrt{2})h^{+}_{\sigma}$ then the matrix elements of the Hubbard
operators in the representation (\ref{99a}) and (\ref{11a})
 are the same between the physical states $|s>, |0 \sigma>$. Moreover
the action of the Hubbard operators in the form (\ref{11a})
on the physical states $|s>,|0 \sigma>$ does
not lead to the unphysical states $|t m>,\
|2 \sigma>$, i.e., the physical states form invariant
subspace over the algebra (\ref{11a}) . The relation
between the Hubbard operators
 $X_{HP}^{ab}$ (\ref{99a}) and $X_{DM}^{ab}$  (\ref{11a})  can
be represented in a form similar to (\ref{3a})
\begin{equation}
\label{12a}
X^{ab}_{HP}=PV^{-1}X^{ab}_{DM}VP
\end{equation}
here V is the generator of the canonical transformation determined
in the physical subspace
\begin{equation}
\label{13a}
V_{ss}=\sqrt{2},\ \ \ V_{\sigma,\sigma^{`}}=\delta_{\sigma,\sigma^{`}},
\ \ \ \ V_{s,\sigma}=V_{\sigma,s}=0
\end{equation}
and P is the projector on physical subspace
\begin{equation}
\label{14a}
P=(1/4)(2-\hat{n})[2-\hat{n}-2\hat{n}(h^{+}{\bf \sigma}h){\bf s}].
\end{equation}
Naturally the operators P and V commute. The partition function
of the Hubbard model can be presented in the form similar to (\ref{5a}),
(\ref{6a})
\begin{equation}
\label{15a}
Z=Tr(P\exp(-\beta H(X^{ab}_{HP}))=Tr(P\exp(-\beta H(X^{ab}_{DM}))
\end{equation}
Since the energies of the two-hole states and the triplet states are
essentially higher , one
can omit the projector P in (\ref{15a}) at the low temperatures
and work with the transformed Hamiltonian . This approximation is valid for
small number of holes so far as the energy of the two-hole
states at the lattice site and the triplet state are situated
approximately at the center of the singlet hole bands.\\

\newpage

\section{Renormalization of the hole energy in an antiferromagnetic state
of the $t-J$ model}
\clequ \

 In this part of the work we will concern ourselves with an antiferromagnetic
state for the  $t-J$  model \cite{be,ka,va} and
will discuss the hole energy.

As was recently proposed by Kane, Lee and Read \cite{ka} (KLR hereafter)
in the limit that the exchange energy $J$ is much less than the
hopping matrix element $t$, the hole spectrum is strongly renormalized
by the interaction with spin excitations.
It was established that the hole can be described by a narrow
quasiparticle band with the quasiparticle residue of an order $J/t$ and
a bandwidth of an order $J$, both for the Neel and RVB groundstates.  Here
we consider the groundstate of the local
spins to be a quantum Neel state.  By applying KLR approach we will show that
our case is rather more complicated than in "holon" model, but the results are
the same.

 An antiferromagnetic state for the Hubbard model was considered
in Ref. \cite{be}. The Hamiltonian of the $ t-J $ model in Ref. \cite{be}
was obtained on the basis of the nonhermitian representation for the
Hubbard operators (\ref{11a})
\begin{eqnarray}
\label{666}
 H_{t-J}=t \sum_{<n,n'>}h^{+}_{n'}[-1/2-(h^{+}_{n}h_{n})+({\bf S_{n}}
{\bf \sigma})] h_{n}+J \sum_{<n,n'>}{\bf S_{n}}{\bf S_{n'}}
\end{eqnarray}
Retaining the notations of \cite{be} we will write the Hamiltonian
of the model.

The Heisenberg part of the Hamiltonian (\ref{666})
expressed through the bose-operators
 $b$ and $c$, connected by $u-v$ transformation with the
primary Dyson-Maleev  bose-operators, which are associated with
two Neel subluttices. It has the form:
\begin{eqnarray}
\label{1a}
&&  H^{s}= \sum_{\bf k}\Omega_{\bf k}(b^{+}_{\bf k}b_{\bf k} +
 c^{+}_{\bf k}c_{\bf k}),
\nonumber\\
&&  \Omega_{\bf k} = zJ(1-\gamma ^{2}_{\bf k})^{1/2},
\nonumber\\
&&  \gamma _{\bf k}=1/2(cos(k_{x}a)+cos(k_{y}a)).
\end{eqnarray}

Next we consider the addition of holes.  The hopping Hamiltonian (\ref{666})
expressed in terms of two-subluttice excitations and holes in lower band is
\begin{eqnarray}
\label{2a}
&&H^{sp1} = 1/2\sum_{\bf k,q} |t_{\bf k}|(U_{\bf q}+ V_{\bf q})
[(b^{+}_{\bf q} + c_{-\bf q})
p^{+}_{\downarrow \bf k-q}p_{\uparrow \bf k} +
(b_{\bf q} + c^{+}_{-\bf q})p^{+}_{\uparrow \bf k+q}
p_{\downarrow \bf k}]
\nonumber\\
&&H^{sp2}=1/2\eta ^{2}\sum_{\bf k,q,k'}|t_{\bf k}|
[p^{+}_{\uparrow \bf k-q}p_{\uparrow \bf k}(-1/2 +V^{2}_{\bf k'}+
\nonumber\\
&&U_{\bf k'}U_{\bf k'+q}b^{+}_{\bf k'+q}b_{\bf k'}+
 V_{\bf k'}V_{\bf k'+\bf q}c^{+}_{\bf k'+\bf q} c_{\bf k'} +
 U_{\bf q-\bf k'}V_{\bf k'} b^{+}_{-\bf k'+\bf q} c^{+}_{\bf k'} +
\nonumber\\
&&V_{-\bf k'}U_{\bf k'-\bf q}c_{-\bf k'}b_{\bf k'-\bf q}+
 p^{+}_{\downarrow \bf k-\bf q}p_{\downarrow \bf k} (b \leftrightarrow c)],
\end{eqnarray}
where $ \eta ^{2}=1/\sqrt{4z}$; $U_{\bf q}=\left( (1+\nu_{\bf q})/
(2\nu_{\bf q})\right)^{1/2}$, $V_{\bf q}=\left( (1-\nu_{\bf q})/
(2\nu_{\bf q})\right)^{1/2}$, $\nu_{\bf q}=(1-\gamma _{\bf q}^{2})^{1/2}$,
$ p^{+}_{\alpha \bf k},p_{\alpha \bf k} $ are holes operators in
the lower hole band.The first two terms in the round brackets
in $ H^{sp2}$ gives in mean-field approximation the primary dispersion
of the holes: $\omega _{\bf k}=|t_{\bf k}|/\sqrt{4z}$, where  $z$ -number
of the nearest  neighbours,  $ t_{\bf k}\equiv zt \gamma _{\bf k}$ .

The terms in (\ref{2a}) with unconserving of the number of the particles
give the two self-energy diagrams at zero temperature corresponding to
emission and absorption one and two magnons.

 So the differences between the "holons", or spinless holes approach
\cite{ka,va} and this model are: (1) the existence of the primary
mean-field dispersion of the hole ,(2) two types of the self-energy
diagrams which strongly renormalized the hole spectrum.

 One can obtain from the Hamiltonian (\ref{2a})
the expression of the self-energy for the hole with particular spin
projection.

It consists of the two contributions containing emission and absorption one
and two magnons.
  Since the model is symmetric over the spin projections
the self-energy in the non-crossing approximation (KLR) has the form:
\begin {eqnarray}
\label{1}
&& \Sigma ({\bf k},\omega )  = \omega ^{0}_{\bf k} +\sum_{\bf q}f_{1}({\bf k,
q}) G({\bf k-q},\omega -E_{\bf q}) +
\nonumber\\
&&\sum_{\bf q,k'}f_{2}
({\bf k,q,k'}) G({\bf k-q},\omega -E_{\bf k'}-E_{\bf k'-\bf q}),
\end{eqnarray}

Where $f_{1}$ and $f_{2}$ - "two-hoppes" functions, which contain
 the information
about different types of coupling of a hole with spin excitations.
 One can easily get from (\ref{2a})
\begin {eqnarray}
\label{2b}
&& f_{1}({\bf k,\bf q}) = (1/4) t_{\bf k}t_{\bf k -\bf q}(U_{\bf q}
+ V_{\bf q})^{2}
\nonumber\\
&&f_{2}({\bf k,\bf k',\bf q}) = zt_{\bf k}t_{\bf k-\bf q}V^{2}_{\bf k'
-\bf q}U^{2}_{\bf k'}.
\end{eqnarray}

In the KLR method we have the selfconsistent integral equation for
the hole propagator:
\begin{eqnarray}
\label{2}
 G({\bf k},\omega) =\frac{1}{\omega -\Sigma ({\bf k},\omega)}.
\end{eqnarray}

Using the dominant pole approximation we write the hole propagator as,
\begin{eqnarray}
\label{3}
G({\bf k},\omega) =\frac{a_{\bf k}}{\omega -\omega _{\bf k}+i\Gamma_{\bf k}}
 + G_{inc},
\end{eqnarray}
where $ \omega _{\bf k}= Re\Sigma ({\bf k},\omega _{\bf k})$,and
$\Gamma_{\bf k} =
 Im\Sigma ({\bf k},\omega _{\bf k})$.
 As pointed out in Ref.\cite{ka} there is a general statement that
$\Gamma_{\bf k} =0$ for all of the low-energy poles.  Since we are
interested in the low energies, we can set $\Gamma_{\bf k} =0$.

 The quasiparticle residue in (\ref{3})
\begin{eqnarray}
\label{4}
 a_{\bf k} =\frac{1}{1- \frac{\partial \Sigma}
{\partial \omega }({\bf k},\omega _{\bf k})}
\end{eqnarray}
may be estimated by omitting the incoherent part in (\ref{3})
\begin{eqnarray}
\label{5}
&& a_{\bf k} \leq [1 +\sum_{\bf q}f_{1}({\bf k,q})\frac{a_{\bf k-q}}
{(\omega _{\bf k}-\omega _{\bf k-q} -E_{\bf q})^{2}}+
\nonumber\\
&&\sum_{\bf q,\bf k'} f_{2} ({\bf k,\bf q,k'})\frac{a_{\bf k-\bf q}}
{(\omega _{\bf k}-\omega _{\bf k-\bf q}
 -E_{\bf k'} -E_{\bf k'-\bf q})^{2}}]^{-1}.
\end{eqnarray}

By applying the scheme of \cite{ka} we will evaluate the integrals in  the
denominator of (\ref{5}) and will show that both integrals are of
 the order $t/J >>1 $.

Near the bottom of the spectrum, where $ \omega_{k} =0 $ and in the
case $ J=0 $  we get the divergence of the integrals at the small $|\bf q|$:
\begin{eqnarray}
\label{6}
&& I_{1} = (t^{2}z^{2}/8\sqrt{2}) \gamma ^{2}_{\bf k^{*}}
\sum_{\bf q}|{\bf q}|\frac{a_{\bf k^{*}}}
{\left[ \frac{q^{2}}{2m^{*}}\right]^{2}}
\nonumber\\
&& I_{2} = (t^{2}z^{3}/2) \gamma ^{2}_{\bf k^{*}}
\sum_{\bf q,k'}\frac{1}{|{\bf k'}||{\bf k'-q}|}
\frac{a_{\bf k^{*}}}{\left[ \frac{q^{2}}{2m^{*}}\right]^{2}},
\end{eqnarray}
where $m^{*}$ is the mass at the bottom of the band. Throughout this work the
distance between lattice sites $a=1$. The both integrals (\ref{6})
are diverged in the dimensions
two and three. Here we will be concerned with only in the
case of $d=2$. In the case of small but finite  $J$ we have
\begin{eqnarray}
\label{7}
I'_{1} \sim t^{2}\sum_{\bf q}|{\bf q}|\frac{a_{\bf k^{*}}}
{\left[ \frac{q^{2}}{2m^{*}}+J|{\bf q}|\right]^{2}}\approx
 t^{2}\frac{m^{*}a_{\bf k^{*}}}{J}.
\end{eqnarray}

Furthermore, as shown in \cite{ka} $m^{*}=m/a_{\bf k^{*}}$ where $m$
depends only on $t$ and $m \sim 1/t$. So
\begin{eqnarray}
\label{8}
I'_{1} \sim \frac{t}{J}.
\end{eqnarray}

The integral $I'_{2}$ has the more complicated form. In the case $d=2$
\begin{eqnarray}
\label{9}
I'_{2} \sim t^{2}\int \int \int d|{\bf q}| \ d|{\bf k'}| \ d\varphi
\frac{|{\bf q}|}{|{\bf k'-q}|}
\frac{a_{\bf k^{*}}}{\left[ \frac{q^{2}}{2m^{*}}+J|{\bf k'}|+J|{\bf k'-q}|
\right]^{2}},
\end{eqnarray}
and its mathematical treatment is unwieldy, but the result is the same
\begin{eqnarray}
\label{10}
I'_{2} \sim t^{2}\frac{m^{*}a_{\bf k^{*}}}{J} \sim \frac{t}{J}.
\end{eqnarray}

thus, the quasiparticle residue $ a_{\bf k} \leq J/t$.

By making physically justified assumption about the behavior of the
imaginary part of the self-energy, one can obtain a rough estimation
for $ a_{\bf k} $ which is in coincides with (\ref {10}). Now this
assumption may be checked.

Using the Kramers-Kroning relation we have for the quasiparticle residue
\begin{eqnarray}
\label{11}
 a_{\bf k} =\left(1 +\int \Gamma({\bf k},y)/(y-\omega _{\bf k})^{2}
dy\right)^{-1}.
\end{eqnarray}
As it was argued by KLR,
 $\Gamma({\bf k},\omega)$  vanishes
like a power of $(\omega-\omega_{\bf k})$ at
$(\omega-\omega_{\bf k}) \ll J$ .
At $(\omega-\omega_{\bf k})\geq J $
we expect that the scattering dominates
and $\Gamma({\bf k},\omega)\simeq t$  . Therefore, the integral in the
 denominator of (\ref {11}) may be cut at $y=J$
\begin{eqnarray}
\label{12}
 a_{\bf k} = \left(1 +\int_{J}^{\infty} y^{-2} dy\right)^{-1}.
\end{eqnarray}

 The behavior of the imaginary part of the self-energy in non-crossing
approximation may be obtained from expressions
\begin{eqnarray}
\label{13}
&& \Gamma({\bf k},\omega) = \frac{1}{\pi}Im \Sigma({\bf k},\omega)=
\Gamma_{1}({\bf k},\omega)+\Gamma_{2}({\bf k},\omega),
\nonumber\\
&& \Gamma_{1}({\bf k},\omega) = \sum_{\bf q}f_{1}({\bf k,q})
A({\bf k-q},\omega -E_{\bf q}),
\nonumber\\
&& \Gamma_{2}({\bf k},\omega) = \sum_{\bf q,k'}f_{2}({\bf k,q,k'})
A({\bf k-q},\omega -E_{\bf k'}-E_{\bf k'-q}),
\end{eqnarray}
where $A({\bf k},\omega )=\frac{1}{\pi}Im \Gamma({\bf k},\omega)$ is the
 spectral function.

For $(\omega-\omega_{\bf k}) \ll J$, the dominant contribution to
 $\Gamma({\bf k},\omega)$ will come from the pole of $A({\bf k},
\omega )$ . Furthermore, the leading contribution will be given by the
very small
$|{\bf q}|$ and $|{\bf k'}|$ . For $\Gamma_{1}({\bf k},\omega)$ we may
write
\begin{eqnarray}
\label{14}
&& \Gamma_{1}({\bf k},\omega) \approx  (t^{2}z^{2}/8\sqrt{2})
 \gamma ^{2}_{\bf k} a_{\bf k}\sum_{\bf q}|{\bf q}|
\delta (\omega-\omega_{\bf k}-J|{\bf q}|) \approx
 \gamma ^{2}_{\bf k}\frac{t^{2} a_{\bf k}}{J}\left(
\frac{\omega-\omega_{\bf k}}{J}\right)^{2}.
\end{eqnarray}
So, $\Gamma_{1}({\bf k},\omega) \sim (\omega-\omega_{\bf k})^{2}$ very near
to the pole. Extrapolating Eqs. (\ref{13}) in the region
$(\omega-\omega_{\bf k}) \sim J$ we get $\Gamma_{1}({\bf k},\omega)\sim t$ .
 For $\Gamma_{2}({\bf k},\omega)$ we obtain
\begin{eqnarray}
\label{15}
&& \Gamma_{2}({\bf k},\omega) =  (t^{2}z^{3}/2) \gamma ^{2}_{\bf k}
\frac{a_{\bf k}}{J} \sum_{\bf q,k'}\frac{1}{|{\bf k'}||{\bf k'-q}|}
\delta (\frac{\omega-\omega_{\bf k}}{J}-|{\bf k'-q}|-|{\bf k'}|) .
\end{eqnarray}
After some mathematical treatment one can get
\begin{eqnarray}
\label{16}
\Gamma_{2}({\bf k},\omega) \approx \frac{t^{2} a_{\bf k}}{J}\left(
\frac{\omega-\omega_{\bf k}}{J}\right)^{2},
\end{eqnarray}
and at $(\omega-\omega_{\bf k}) \sim J \  \Rightarrow  \
\Gamma({\bf k},\omega)\sim t$ .

 Thus, we have demonstrated the application of the approach of KLR method to
this model
and have obtained the similar results, i.e. strong renormalization of a hole
spectrum, narrow quasiparticle band with the quasiparticle residue $J/t$
and the bandwidth $J$ .

\section{The functional integral for the strongly correlated system}
\clequ \

Several types of functional integrals can be generated by the
representation (\ref{99a}) for the Hubbard operators.  We shall consider two
types of such representation. The first can be named "constraint"
representation.  The second can be named as "compensation" representation.

For the first type of a functional integral we shall use some representation
for spin$-\frac{1}{2}$ operators
\begin{eqnarray}
\label{a18}
&&{\bf S}=1/2(b^{+}{\bf \sigma}b), \ \  A_{s}=(b^{+}b)-1 =0
\end{eqnarray}
here $ b^{+}_{\alpha },b_{\alpha }$ for $\alpha  =\uparrow ,\downarrow  $
are creation and annihilation operators which can be boson or fermion
operators, $A_{s}$ is the spin antiprojector operator which is equal
to zero on the
physical subspace and it is equal to $ n=-1,1,2,3 $ on the unphysical
subspace. The condition  $ A_{s} =0 $ is usually named a constraint, but
it is very special type of a constraint. The antiprojector operator  $A_{h} $
can also be introduce for the representation (\ref{99a}) for the
Hubbard operators:
\begin{eqnarray}
\label{a19}
&& A_{h} = (1/4)(3\hat{n}^{h} - 2\hat{d}^{h}+ 2{\bf S}
h^{+}{\bf \sigma }h).
\end{eqnarray}

This antiprojector is equal to unity on the two-hole state and on the
one-hole triplet states. It is equal to zero on the hole vacuum
 state and  on the one-hole singlet state.
The introduction of the antiprojector $A$ is motivated by an identity
connecting it with the projector $P$:
\begin{eqnarray}
\label{a20}
&& P= (1/2\pi )\int_{0}^{2\pi }\exp(i\lambda A)d\lambda   .
\end{eqnarray}
The identity (\ref{a20}) gives us possibility to construct the functional
integral
for any dynamical system which can be expressed in terms of canonical
variables and act on a limiting part of the Gilbert space which is defined by
the condition  $ A=0 $.
In our case  of a strongly correlated system we have the following
representation for the partition  function or the generating functional of
the Green functions
\begin{eqnarray}
\label{a21}
&& Z(y_{l}^{ab}) = \int\int  \exp( S + i\int_{0}^{\beta}
 d\tau  \sum_{l}(\lambda_{sl}
((b^{+}_{l}b_{l})-1) +
\nonumber\\
&&\lambda_{Hl}(3\hat{n}_{l}^{h}-2\hat{d}_{l}^{h} +  2{\bf S}_{l}
h_{l}^{+}{\bf \sigma}h_{l})) \delta \left(\frac{\partial\lambda_{sl}
(\tau )}{\partial\tau }\right) \delta \left(\frac{\partial\lambda _{Hl}
(\tau)}{\partial\tau }\right)
\nonumber\\
&& \prod_{l,\tau ,\alpha }Dh^{+}_{l\alpha }(\tau ) Dh_{l\alpha} (\tau)
Db^{+}_{l\alpha }(\tau ) Db_{l\alpha  }(\tau )D\lambda _{sl}(\tau )
D\lambda_{Hl}(\tau ),
\nonumber\\
&& S=\int_{0}^{\beta} d\tau \sum_{l}(h^{+}_{l}\dot{h}_{l}+
b^{+}_{l}\dot{b}_{l} -H(h^{+}_{l} h_{l},{\bf S}_{l}(b^{+}_{l},b_{l})) +
\nonumber\\
&& X^{ab}_{l}(h^{+}_{l},h_{l}, {\bf S}_{l}(b^{+}_{l},b_{l}))y_{l}^{ab})
\end{eqnarray}
Here S is the action of  a
system;  $H( h^{+},h, {\bf S}(b^{+},b))$ is the Hamiltonian
of a strongly correlated system; l is the lattice site index;
$\lambda_{s},\lambda _{H}$
are the Lagrange multipliers which do not depend on the
temperature time $\tau $ what is ensured due to the presence of $\delta $-
functions. Because the spin subsystem possess the gauge invariance the
$\delta (\partial\lambda_{s}(\tau )/\partial\tau )$ fixes the
gauge.

 Another possibility to construct the functional integral
is connected with the idea of compensation of an unphysical contributions
suggested in the work by Porov and Fedotov \cite{pf}. They use the
equality to zero of the
Hamiltonian on the unphysical states.  Let us add to the Hamiltonian some
addition $ \delta H $ which is equal to zero on the
physical states and possess
the properties
\begin{eqnarray}
\label{a22}
Tr(\exp(-\beta \delta H ))_{HP}=
0 , \ \ [ H,\delta H ] =0
\end{eqnarray}
Summation in  (\ref{a22}) is produced
over the unphysical states only . Then the total Hamiltonian  $ H_{t}=H
+\delta H $ can be used in the total Gilbert space
without any constraints for
computation of the partition function.  The functional in this case has
simple form:
\begin{eqnarray}
\label{a23}
Z(y_{ab}) = \int \exp(S_{\tau }) \prod_{l,\tau ,\alpha }
Dh^{+}_{l\alpha }(\tau ) Dh_{l\alpha } (\tau )Db^{+}_{l\alpha }(\tau )
 Db_{l\alpha }(\tau ),
\end{eqnarray}
here $b^{+}_{l\alpha }, b_{l\alpha }$ are the Fermi fields and the addition
to the action comparing with  (\ref{a21}) has form:
\begin{eqnarray}
\label{a24}
\delta S= -\beta\delta  H = 1/2\sum_{l}[i\pi (b^{+}_{l}b_{l} -1) -(i\pi
+\ln (3/2))(3h^{+}_{l}h_{l}+ 2{\bf S}_{l}h^{+}_{l}{\bf \sigma}
h_{l})]
\end{eqnarray}
This addition to the action is nonhermitian. In fact we can easily
 construct a two-parametric form of $\delta S $.

\section{ The representation of the total Hubbard algebra}
\clequ \

In some physical situations for example when  the metal-insulator
transition is studied
it can be necessary to describe holes in the lower Hubbard band
and electrons in the upper Hubbard band
simultaneously.  In this case it will be convenient to have representation
that describes these electrons and holes as the different degrees of
freedom simultaneously.  For realization of this possibility we shall
consider the additional Gilbert space consisting of the states of
spin$-\frac{1}{2}$
 and two types of fermion with spin$-\frac{1}{2}$
 for describing electrons in the
upper band and holes in the lower bonds.  The mapping of the physical
electrons states $|0>, |\uparrow >,\downarrow >, |2>$ into states
of the additional Gilbert space is following:
\begin{eqnarray}
\label{a25}
&& |\uparrow > \ \Rightarrow \ |s\uparrow ) ,\ \ \ \ \  \ \ \ \
|\downarrow > \ \Rightarrow \ |s\downarrow )
\nonumber\\
&& |0> \ \Rightarrow \  (1/\sqrt{2}) [h^{+}_{\uparrow }|s\downarrow )
-h^{+}_{\downarrow }|s\uparrow )],
\nonumber\\
&&|2> \ \Rightarrow \ (1/\sqrt{2}) [e^{+}_{\uparrow }|s\downarrow ) -
e^{+}_{\downarrow }|s\uparrow )].
\end{eqnarray}
Here $ |s\alpha )$ for $\alpha  =\uparrow ,\downarrow $ are the states
of the spin$-\frac{1}{2}$;
 $ h^{+}_{\alpha },h_{\alpha }, e^{+}_{\alpha }, e_{\alpha }$
are the creation and annihilation the Fermi
operators of holes in the lower band and electrons in the upper band.  The
physical states are singlets formed from the local spin$-\frac{1}{2}$
and the hole (the electron) spin$-\frac{1}{2}$.
One can get a representation of the total Hubbard
algebra based on the described mapping.  The representation of the lower
subalgebra has form:
\begin{eqnarray}
\label{a26}
&& X^{0\alpha } = 2\alpha \sqrt{2} (h^{+}\hat{S})_{-\alpha}(1-n^{h})P_{e}
\nonumber\\
&& X^{\alpha 0}=2\alpha\sqrt{2}  P_{e}(1-n^{h}) (\hat{S} h)_{-\alpha }
\nonumber\\
&& X^{00}= (1/4) (n^{h}-2d^{h}-2t^{h}) P_{e}.
\end{eqnarray}

The representation of the upper subalgebra is quite similar
\begin{eqnarray}
\label{a27}
&& X^{2\alpha }= 2\alpha\sqrt{2} (e^{+}\hat{S})_{-\alpha}(1-n^{e})P_{h}
\nonumber\\
&& X^{2\alpha }= 2\alpha\sqrt{2}  P_{h}(1 - n^{e}) (\hat{S} e)_{-\alpha }
\nonumber\\
&& X^{22}= (1/4) (n^{e}-2d^{e}-2t^{e})P_{h}.
\end{eqnarray}
Here we use the following short notation:
\begin{eqnarray}
\label{a28}
&& \hat{S}= (1/4) (1-2{\bf S}{\bf \sigma }), \ \ \
P_{c}=1-n^{c}+d^{c},
\nonumber\\
&& n^{c}= (c^{+}c) , \ \ \ \ \  d^{c}= c^{+}_{\uparrow }c_{\uparrow }
c^{+}_{\downarrow }c_{\downarrow},
\nonumber\\
&& t^{c}= {\bf S}c^{+}{\bf \sigma }c ,\ \ \ \  for \ \ \ c: =h,e \ .
\end{eqnarray}
The representation of the middle subalgebra has form:
\begin{eqnarray}
\label{a29}
&& N=X^{\uparrow \uparrow }+X^{\downarrow \downarrow}  =P_{e}P_{h}
\nonumber\\
&& {\bf S}=(1/2){\bf \sigma}_{\alpha \beta }X^{\beta \alpha}=
{\bf s}P_{e}P_{h},
\end{eqnarray}
here ${\bf s}$ is  operator of the local spin. Using this representation for
the Hubbard operators we can easily construct the functional integral and the
diagram technique.  Such type representation can be also useful in
variational approach for the construction of the explicit form of
interaction the charge carriers with the local spins.

\section{ Acknowlegment}

Author is helpful S.V.Maleev for useful conversation.  This work was
supported partly by the Counsel on Superconductivity of Russian Academy of
Sciences ,Grant No.90214.

\end{document}